# Theory of Azimuthally Propagating Electromagnetic Waves in Cylindrical Cavities

Mustafa Bakr, and Smain Amari

*Abstract*— The paper presents a detailed study of azimuthally propagating electromagnetic waves in cylindrical metallic cavities with circular cross section. Dispersion characteristics of these waves are determined from Maxwell's equations. Solutions are grouped into branches that account for all known results that are obtained from axial propagation. It is reported that the lowest TE mode starts propagating in the azimuthal direction at a frequency that depends only on the height of the cavity and may be much lower than the cutoff the $TE_{11}$ mode in the axial direction. Universal curves allowing the determination of resonant frequencies and field distribution of TE and TM modes in circular cavities containing wedges of arbitrary angles and baffles with no additional computation. It is shown that the frequency dependence of the propagation constant of a given branch determines all the resonant frequencies of the branch for arbitrary boundary conditions in azimuthal direction. It is argued that propagation-based models, when applicable, are more accurate than resonance-based models. The lowest TE branch starts at a non-physical resonance. Applications to microwave dual-mode filter design are discussed briefly.

## I. Introduction

A survey of the vast literature on microwave filters, and microwave bandpass filters in particular, shows that there are two distinct central ideas behind their modelling and design. The first idea is based on the use of the phenomenon of propagation and the second on the phenomenon of resonance.

The seminal work of Cohn on direct-coupled resonator filters is arguably the best example of propagation-based models [1]. Other works on transmission line based filters, through Richards transformation, fall within this category [2]. A convincing example of the success and accuracy of transmission-line-based models is the solution of the interdigital filter problem with arbitrary bandwidth by Wenzel [3]. A limitation of this model is its applicability only to structures which contains uniform sections, or sub-sections, where propagation is possible.

The most influential investigations of models based on the phenomenon of resonance are possibly those of Slater, Schelkunoff, Kurokawa and Reiter [4-7]. In this approach, the electromagnetic field inside a cavity is represented by an infinite series over the eigen-modes of the cavity. Each resonance is dominant around its resonant frequency especially when it is coupled weakly to the ports. This is the case of narrow-band filters. As the bandwidth increases, resonances away from the passband become important and cannot be ignored. Unfortunately, resonance-based models provide no mechanism to take these 'spurious' resonances into account. As a consequence, extensive optimization is often required to design resonance-based microwave filters especially as the bandwidth increases. On the positive side, resonance-based models are applicable to any coupled-resonator system regardless of the shape of the resonator.

Propagation of EM waves along a TEM line is well understood. A lossless TEM line placed along the z-axis supports waves of the form $e^{\mp j\beta z}$ where β is the propagation or the phase constant. It is also known that β is a linear function of the frequency $\beta = \frac{\omega}{v_p}$ where ω is the angular frequency and $v_p$ the phase velocity.

Similarly, the $H_z$ component of the electromagnetic field of a TE-to-z mode in a circular wave guide, filled with a material with parameters ε and μ, is of the form $J_\nu(k_c\rho)e^{\mp j\nu\theta}e^{\mp jk_z z}$ with $k_z^2 = \omega^2\mu\epsilon - k_c^2$. Here $J_\nu(k_c\rho)$ is Bessel function of the first kind of order ν, $k_z$ is the propagation constant along the axis of the waveguide and $k_c$ is the cut-off wavenumber of the mode. Let us now assume that perfectly conducting plates are placed at z=0 and z=h. The boundary conditions at these plates reduce the possible values of $k_z$ to the discrete set $k_z = \frac{p\pi}{h}, p = 1,2,3 \ldots$. For the lowest mode, p=1, we then have

$$H_z = J_\nu(k_c\rho)e^{\mp j\nu\theta} \sin\left(\frac{\pi}{h}z\right) \quad (1)$$

$$\omega^2\mu\epsilon = k_c^2 + \left(\frac{\pi}{h}\right)^2 \quad (2)$$

If we include the harmonic time dependence in equation (1), we get

$$H_z(\rho,\theta,z,t) = J_\nu(k_c\rho)\cos(\omega t \mp \nu\theta)\sin\left(\frac{\pi}{h}z\right) \quad (3)$$

Two important points to note about this expression: a) it represents waves that are propagating in the $\theta$ direction with "azimuthal propagation constant" ν (or -ν), and b) the value of ν is not determined yet since we have not imposed the boundary conditions in the azimuthal direction. However, for equation (3) to represent an electromagnetic field, it must satisfy Maxwell's equations along with the boundary conditions. It is straightforward to show that it satisfies the wave or Helmholtz equation. It satisfies the boundary conditions at z=0 and z=h. The boundary conditions in the radial direction are yet to be enforced. It will be shown that

Mustafa Bakr is with the Department of Physics, The University of Oxford, Oxford, OX 3PU, UK (e-mail: mustafa.bakr@physics.ox.ac.uk)

Smain Amari is with The Royal Military College of Canada, Kingston, ON K7K 7B4, Canada (e-mail: smain.amari@rmc.ca).

enforcing these boundary conditions imposes a relationship between the azimuthal propagation constant ν and the angular frequency ω. The result is the dispersion relation of these waves.

The main goal of this study is to determine how the azimuthal propagation constant ν varies with frequency in a circular cavity or in azimuthally uniform sections of a circular cavity. This variation plays a central role in developing propagation-based models of sophisticated classes of dual-mode filters in side-coupled circular cavities and other microwave components. Examples of dual-mode filters in side-coupled cavities based on azimuthal propagation are reported in [8-9]

The key findings reported in this paper are:

1. When applicable, propagation-based models are vastly superior to resonance-based models.
2. The 'azimuthal propagation constants" of TE modes in a uniform lossless cylindrical cavity are grouped in separate branches.
3. Each TE branch starts from a root of $J_1(x)=0$ when ν=0.
4. The lowest TE branch starts from $x=0$ when ν=0.
5. Modes belonging to different TE branches are orthogonal for the same value of ν.
6. Resonant solutions belonging to the same TE branch are orthogonal.
7. The propagation constants of the TM modes in a uniform lossless cylindrical cavity are grouped in separate branches.
8. Each TM branch starts from a root of $J_0(x)=0$ when ν=0.
9. The lowest TM branch starts at x=2.4050 when ν=0.
10. Modes belonging to different TM branches are orthogonal for the same value of ν.
11. Resonant solutions belonging to the same TM branch are orthogonal.
12. Modes belonging to any TE branch are orthogonal to modes belonging to any TM branch.
13. The "azimuthal propagation constants" are purely imaginary below their respective cut-off frequencies.

We start by discussing two examples that illustrate the 'superiority' of propagation-based models, when they are applicable, over resonance-based models.

## II. PROPAGATION VERSUS RESONANCE

### A. Shorted TEM line

Consider a lossless TEM transmission line of physical length $l$. The line is terminated by a short circuit at one end and fed by an ideal sinusoidal voltage source $V_i$ of angular frequency ω at the other end as shown in Figure 1. We assume that the inductance and the capacitance per unit length of the line are L and C. We would like to determine the input admittance of this line.

From undergraduate textbooks [], we know that the input admittance is given by

$$Y_{in} = -\frac{jY_o}{\tan(\beta l)}, Y_o = \sqrt{\frac{C}{L}}, \beta = \sqrt{LC}\omega \quad (4)$$

Here, $Y_o$ is the characteristic admittance and β the propagation constant. This is indeed a very simple and accurate formula over the entire frequency range where only the TEM mode is propagating. The key thing to note here is that this formula is based on the superposition of incident and reflected TEM waves which are propagating on the line.

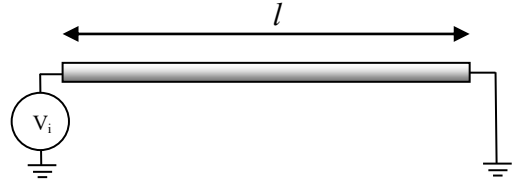

Figure 1. TEM transmission line terminated in a short circuit and fed by an ideal voltage source $V_i$.

We now derive the same quantity by using the concept of resonance. We know that the resonant frequencies are the poles of the input admittance given by equation (4). These are given by

$$\tan(\sqrt{LC}l\omega) = 0 \Rightarrow \omega_n = \frac{n\pi}{\sqrt{LC}l} = n\omega_o, n = 0,1,2\ldots \quad (5)$$

Here, we introduced the characteristic angular frequency $\omega_o = \frac{\pi}{\sqrt{LC}l}$. To express the input admittance in terms of these resonant modes, we can follow one of two options. One option consists in starting from the differential equations of the voltage and current on the transmission line followed by a Fourier series expansion of the current and the voltage on the line. The details are given by Schelkunoff and are not repeated here [11, pp. 272-276]. A second option consists in using the partial fraction expansion of the function cot(x), namely [12, p.44]

$$\cot(x) = \frac{1}{x} + \sum_{n=1}^{\infty} \frac{2x}{x^2 - n^2\pi^2} \quad (6)$$

By using equation (6) in (4), we get

$$Y_{in} = -jY_o\left[\frac{1}{\beta l} + \sum_{n=1}^{\infty} \frac{2\beta l}{(\beta l)^2 - n^2\pi^2}\right] \quad (7)$$

In terms of the characteristic angular frequency $\omega_o$ and the characteristic admittance $Y_o$ this takes the form

$$Y_{in} = -\frac{jY_o}{\pi}\left[\frac{\omega_o}{\omega} + \sum_{n=1}^{\infty} \frac{2\omega_o\omega}{\omega^2 - \omega_n^2}\right] \quad (8)$$

Equations (4) and (8) give the same quantity. Although the sum in equation (8) can be computed, this is rarely the case in resonance-based models. In a resonance-based model, most often only one resonance is taken into account and the remaining ones are ignored. The advantage of the propagation-based model is that it includes *all* the resonances that are brought about by the relevant propagating mode. This is not surprising since propagation "knows" about the local behaviour of the electromagnetic field as well as the boundary conditions that set up the resonances whereas the resonances are determined mainly by the global boundary conditions. What disadvantages a resonance-based model is its inability to include more than few resonances in actual design. Valuable information on the behaviour of the system may be lost when resonances are given the prominent role over propagation.

It is also important to note that if the boundary conditions change, such as changing the length of the line, or terminating it with an open circuit, the resonances are still obtained directly from the propagation constant.

### B. Micrsotrip transmission line ring

Consider the microstrip line ring shown in Figure 2. The microstrip line is assumed narrow enough and the mean radius R sufficiently large for the line to be characterized by a an effective propagation constant of the form

$$\beta_{eff} = \frac{\omega}{v_{eff}} \quad (9)$$

The current and voltage waves propagating around the ring depend on the arc length s = Rθ and θ as

$$e^{\mp j\beta_{eff}s} = e^{\mp j\beta_{eff}R\theta} = e^{\mp j(\beta_{eff}R)\theta} = e^{\mp j\nu\theta} \quad (10)$$

Here, we introduced the frequency dependent parameter ν(ω) by

$$\nu(\omega) = \beta_{eff}R = \frac{R}{v_{eff}}\omega \quad (11)$$

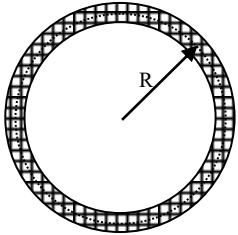

Figure 2. A microstrip transmission line ring of mean radius R.

It is important to note that for this mode, ν(ω) is a continuous linear function of frequency at *any* frequency as equation (11) clearly shows. This is a local property of the line. We call the parameter ν the "azimuthal propagation constant".

If we now apply the boundary condition that the voltage or current on the ring be single-valued, we get the possible values of ν,

$$\nu = n = 0,1,2... \quad (12)$$

This result seems to imply that ν can assume only integer values. However, from equation (11) we know that ν is a *continuous linear* function of frequency. In fact, what equation (12) gives are the values of the azimuthal propagation constant *at* the resonant frequencies of the ring which are given by

$$\omega_{rn} = n\frac{v_{eff}}{R}, n = 0,1,2,... \quad (13)$$

In terms of the guided wavelength $\lambda_g$ this is equivalent to

$$2\pi R = n\lambda_g, n = 0,1,2.. \quad (14)$$

A more rigorous way of writing equation (12) is

$$\nu(\omega_{rn}) = n = 0,1,2,,3... \quad (15)$$

This condition contains no information on how the azimuthal propagation constant depends on frequency other than the fact that it is equal to an integer at the resonant frequencies. If the frequency is equal to one of the resonant frequencies then the mode with that resonant frequency can exist *alone* since it satisfies all the boundary conditions by itself. However, at any frequency that is not equal to any of the resonant frequencies, potentially all the resonant modes must be excited in order to satisfy the boundary conditions. It will then become necessary to include all of them, or at least all those resonances that contribute significantly, in order to describe the frequency response of the structure. Unfortunately, there is no known simple resonance-based method to include all these resonances except for simple cases. On the other hand, knowing the frequency dependence of the propagation constant that generates the resonance allows us to include all the resonances in a systematic way.

It is obvious from this analysis that resonance-based models contain only part of the useful information that is provided in the frequency dependence of the propagation constant. The phenomenon of resonance only samples the propagation constant at the resonant frequencies. The boundary conditions act as a sampling or filtering mechanism of the propagation constant.

Obviously, if the transmission medium supports more than one mode of propagation, the phenomenon of resonance samples the propagation constant of each mode separately. Each propagation constant defines its own branch from which the resonant frequencies are sampled. Again, information on the frequency dependence of the propagation constants of the individual branches is lost.

Let us now go back to the main structure considered in this paper, i.e, cylindrical cavities with circular cross section. We know that the fields of the modes in a circular cavity vary with the azimuthal angle θ in exactly the same way as the waves on a microstrip ring, or

$$e^{\mp j\nu\theta} \quad (16)$$

For the EM field to be single-valued we again end up with the condition (12) with no further information about how the azimuthal propagation constant ν varies with frequency. The central goal in this paper is to derive the frequency dependence of the propagation constants of azimuthally propagating modes and their field distributions. The problem can be stated through the following thought experiment or 'Gedankenexperiment'.

## III. STATEMENT OF THE PROBLEM

We consider a cylindrical cavity of height $h$ with a circular cross section of radius $a_0$. The cavity is assumed filled with a lossless dielectric material with parameters $\varepsilon$ and $\mu$. A cylindrical system of coordinates $(\rho,\theta,z)$ is used.

Let us assume that a mechanism to launch electromagnetic waves in the azimuthal direction at an angular frequency $\omega$ is located at $\theta_0=0$. We also assume that the volume between the planes at $\theta_1=\pi$ and $\theta_2=3\pi/2$ is filled with a perfectly absorbing material (black body). No waves can be reflected from this material. The structure is uniform over the total height of the cavity. The cross section of the set up is shown in Figure 3.

We now place an observer on the circumference at $\theta_{ob}=\pi/2$. Because of the absorbing material, the observer will only detect waves that are propagating in the counter-clockwise direction. The observer is tasked with describing the waves as they move past his/her position at $\theta_{ob}=\pi/2$. In particular, we would like information on the propagation constant of these azimuthally propagating waves, their dispersion relations and field distributions.

Obviously, we cannot rely on resonance to extract the desired information since resonance only samples the propagation constant at the resonant frequencies. We need to look at the problem from a different point of view to extract the propagation constant as a function of frequency, or the dispersion relation of the modes.

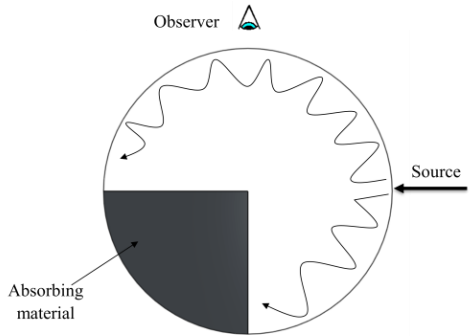

Figure 3. Cross section of structure for thought experiment on azimuthally propagating waves. The observer sees only waves that are propagating counter-clockwise from the source towards the absorbing material. The radius of the cavity is $a_0$ and its height is h.

## IV. A DIFFERENT POINT OF VIEW

The key idea to determine the variation of the azimuthal propagation constant $\nu(\omega)$ with frequency is to recall that a propagation constant is determined by imposing the boundary conditions *only* in the plane that is transverse to the direction of propagation. For example, to determine the constant of propagation in the z-direction in a rectangular waveguide, the boundary conditions are imposed *only* in the xy plane. Once the boundary conditions are imposed in the transverse plane, they are automatically satisfied along the direction of propagation as long as the structure remains uniform in that direction. Similarly, for the cylindrical waveguide with its axis along the z-axis, to determine the axial propagation constant, boundary conditions are imposed only on $\rho$ and $\theta$. Since we are interested in the azimuthal propagation constant, we impose the boundary conditions only in the transverse plane, or equivalently only on $\rho$ and z.

It is well known that the structure under consideration supports modes that can be divided into Transverse Electric (TE) modes and Transverse Magnetic (TM) modes. Here, the transversality is still with respect to the z-axis. We first consider the Transverse Electric (TE) modes.

### 1) TE Modes

The transverse components of the electromagnetic field inside the cavity can be determined from the axial component $H_z$ which is of the form

$$H_z(\rho,\theta,z) = J_\nu(k_c\rho)e^{-j\nu\theta}[Ae^{-jk_z z} + Be^{jk_z z}] \quad (17)$$

Here, A and B are arbitrary constants and $J_\nu(x)$ is the Bessel function of the first kind of order $\nu$. The parameters $k_c$ and $k_z$ are related by

$$k_c^2 + k_z^2 = k^2 = \omega^2\mu\epsilon \quad (18)$$

The boundary conditions at z=0 and z=h on $H_z$, lead to the conditions $A+B = 0$ and

$$k_z = p\frac{\pi}{h}, p = 1,2,3\ldots \quad (19)$$

With $k_z = \frac{p\pi}{h}$, equation (18) becomes

$$k_c^2 = \omega^2\mu\epsilon - \left(\frac{p\pi}{h}\right)^2 \quad (20)$$

The boundary condition in the radial direction leads to

$$J_\nu'(k_c a_0) = J_\nu'\left(a_0\sqrt{\omega^2\mu\varepsilon - \left(\frac{p\pi}{h}\right)^2}\right) \quad (21)$$

Here, $J_\nu'(x)$ is the derivative of the Bessel function with respect to its argument. Let us examine this equation more closely. If the frequency $\omega$ and the value of $p$ are given, the argument in equation (21) is completely determined. The only unknown quantity in the transcendental equation (21) is the order of the Bessel function $\nu$ which is nothing other than the azimuthal propagation constant. This exactly the quantity we are trying to determine. The process is now clear: for a given value of the frequency, determine the order of the Bessel function $\nu$ such that the quantity $a_0\sqrt{\omega^2\mu\varepsilon - \left(\frac{p\pi}{h}\right)^2}$ is a solution of equation (21). A sweep over frequency generates the desired relationship between ν and ω (for a given value of p).

Although this procedure works, it is not universal since the solution depends on the dimensions of the cavity and the frequency. Each time the dimensions change, the procedure would need to be repeated. A universal solution consists in determining how the roots $x'$ of the equation $J'_\nu(x') = 0$ vary with ν as an intermediary step.

Let us call $x'_{\nu n}$ the $n^{th}$ root of the following equation

$$J'_\nu(x'_{\nu n}) = 0 \quad (22)$$

The index $n$ denotes the order of the root and is used to reflect the fact that Bessel functions of the first kind have an infinite number of roots. The smallest root of equation (22) is denoted by $x'_{\nu 1}$, the second root by $x'_{\nu 2}$ and so on. As the azimuthal propagation constant ν varies, the root $x'_{\nu n}$ varies too. By sweeping from ν=0 to any desired value, the variation of the root $x'_{\nu n}$ as a function of ν is determined for any $n$. Once the root $x'_{\nu n}$ is known as a function of ν, the relationship between the frequency ω and the azimuthal propagation constant ν is given by the equation

$$\frac{x'_{\nu n}}{a_0} = \sqrt{\omega^2 \varepsilon \mu - \left(\frac{p\pi}{h}\right)^2}, p = 1,2,3,.. \quad (23)$$

Equations (22) and (23) yield the dispersion relation of the azimuthal TE modes.

The dispersion relation will have a separate branch for each value of the index $n$, the order the root of equation (23). One way to classify these branches is to consider their values for ν=0 where the azimuthally propagating mode starts propagating. However, when ν=0, equation (22) becomes

$$J'_0(x'_{0n}) = -J_1(x'_{0n}) = 0 \quad (24)$$

This equation states that from each zero of the Bessel function $J_1(x)$, starts one branch. One of the most surprising results of this analysis is the appearance of a branch that starts from zero since $x=0$ is indeed a solution of equation (24), or $J_1(0)=0$. Some of the properties of this unusual branch will be discussed later.

At this point, the components of the electromagnetic field of the TE modes are given by (p=1,2,3.., n=1,2,3,..)

$$E_\rho = -A \frac{\omega \mu \nu a_0^2}{(x'_{\nu n})^2 \rho} J_\nu \left(\frac{x'_{\nu n}}{a_0}\rho\right) e^{-j\nu\theta} \sin\left(\frac{p\pi}{h}z\right) \quad (25.a)$$

$$E_\theta = A \frac{j\omega\mu a_0}{x'_{\nu n}} J'_\nu \left(\frac{x'_{\nu n}}{a_0}\rho\right) e^{-j\nu\theta} \sin\left(\frac{p\pi}{h}z\right) \quad (25.b)$$

$$E_z = 0 \quad (25.c)$$

$$H_\rho = A \frac{p\pi a_0}{x'_{\nu n}h} J'_\nu \left(\frac{x'_{\nu n}}{a_0}\rho\right) e^{-j\nu\theta} \cos\left(\frac{p\pi}{h}z\right) \quad (25.d)$$

$$H_\theta = -A \frac{jp\pi\nu a_0^2}{(x'_{\nu n})^2 h\rho} J_\nu \left(\frac{x'_{\nu n}}{a_0}\rho\right) e^{-j\nu\theta} \cos\left(\frac{p\pi}{h}z\right) \quad (25.e)$$

$$H_z = A J_\nu \left(\frac{x'_{\nu n}}{a_0}\rho\right) e^{-j\nu\theta} \sin\left(\frac{p\pi}{h}z\right) \quad (25.f)$$

If we now add the time dependence $e^{j\omega t}$, we see clearly that these local waves are propagating the azimuthal direction. For example, assuming that $A$ is real, the axial component becomes

$$H_z(\rho,\theta,z,t) = A J_\nu \left(\frac{x'_{\nu n}}{a_0}\rho\right) \sin\left(\frac{p\pi}{h}z\right) \cos(\omega t - \nu\theta) \quad (26)$$

This equation unambiguously shows that this is a wave propagating in the positive azimuthal direction for ν positive. The "phase velocity" of this wave is

$$v_p = \frac{\omega}{\nu} \quad (27)$$

$$\vec{P}_\theta = \frac{1}{2} Re[\vec{E} \times \vec{H}^*] = \frac{\omega \mu \nu a_0^2}{2(x'_{\nu n})^2 \rho} \left| A J_\nu \left(\frac{x'_{\nu n}}{a_0}\rho\right) \sin\left(\frac{p\pi}{h}z\right) \right|^2 \hat{u}_\theta \quad (28)$$

This equation shows that the average power flow is in the azimuthal direction. It also shows that changing the sign of the azimuthal propagation constant ν reverses the direction of the power flow as expected.

Finally, the wave impedance of these waves in the azimuthal direction is given by

$$\frac{E_\rho}{H_z} = Z_{TE}(\omega) = \frac{\omega \mu \nu a_0^2}{(x'_{\nu n})^2 \rho} \quad (29)$$

Not surprisingly, it depends on the curvature of the cavity through the radial distance ρ and the radius of the cavity $a_0$ and naturally on the frequency ω and the azimuthal propagation constant ν.

*2) TM modes*

The analysis of TM modes is similar to the case of the TE modes. The components of the electromagnetic field are determined from the axial component $E_z$ which takes the form

$$E_z(\rho,\theta,z) = A J_\nu(k_c \rho) e^{-j\nu\theta} \cos\left(\frac{p\pi z}{h}\right), p = 0,1,2.. \quad (30)$$

The dispersion relations of the different branches are found to be determined from the transcendental equation

$$J_\nu(x_{\nu n}) = 0 \quad (31)$$

Once the roots $x_{\nu n}$ are determined as functions of ν, the variation of the azimuthal propagation constant ν of the $n^{th}$ branch (n=1,2,3..) of the TM modes are given by the equation

$$\frac{x_{\nu n}}{a_0} = \sqrt{\omega^2 \varepsilon \mu - \left(\frac{p\pi}{h}\right)^2}, p = 0,1,2.. \quad (32)$$

The dispersion branches of the TM modes can be classified according to their values at ν=0. In this limit, the transcendental equation (31) becomes

$$J_0(x_{0n}) = 0, n = 1,2,3... \quad (33)$$

Similarly to the TE case, from each of the zeros of the Bessel function $J_0(x)$ starts one dispersion branch for TM modes. However, as opposed to the TE case, there is no branch starting at zero since the smallest root of $J_0(x)=0$ is 2.404826 which is the starting point of lowest TM branch. Once the propagation constant of the different branches are known as a function of frequency, the resonant frequencies can be directly obtained by sampling these branches at the relevant values of ν that are imposed by the boundary condition on the azimuthal variable θ.

The components of the electromagnetic field of the TM modes are given by (p=0,1,2,.., & n=1,2,3,..)

$$E_\rho = \frac{p\pi a_0}{x_{vn}h} A J'_v\left(\frac{x_{vn}}{a_0}\rho\right) e^{-jv\theta} \sin\left(\frac{p\pi z}{h}\right) \quad (34.a)$$

$$E_\theta = \frac{-jva_0^2}{x_{vn}^2}\frac{p\pi}{h} A J_v\left(\frac{x_{vn}}{a_0}\rho\right) e^{-jv\theta} \sin\left(\frac{p\pi z}{h}\right) \quad (34.b)$$

$$E_z = A J_v\left(\frac{x_{vn}}{a_0}\rho\right) e^{-jv\theta} \cos\left(\frac{p\pi z}{h}\right) \quad (34.c)$$

$$H_\rho = A \frac{\omega\epsilon v\, a_0^2}{x_{vn}^2 \rho} J_v\left(\frac{x_{vn}}{a_0}\rho\right) e^{-jv\theta} \cos\left(\frac{p\pi}{h}z\right) \quad (34.d)$$

$$H_\theta = \frac{-j\omega\epsilon a_0}{x_{vn}} A J'_v\left(\frac{x_{vn}}{a_0}\rho\right) e^{-jv\theta} \cos\left(\frac{p\pi z}{h}\right) \quad (34.e)$$

$$H_z = 0 \quad (34.f)$$

If we now add the time dependence $e^{j\omega t}$, we see clearly that these waves are propagating the azimuthal direction. For example, assuming that $A$ is real, the axial component becomes

$$E_z(\rho,\theta,z,t) = A J_v\left(\frac{x_{vn}}{a_0}\rho\right) \cos\left(\frac{p\pi}{h}z\right) \cos(\omega t - v\theta) \quad (35)$$

This equation unambiguously shows that this is a wave propagating in the positive azimuthal direction for ν positive. The "phase velocity" of this wave is

$$v_p = \frac{\omega}{v} \quad (36)$$

To provide another confirmation that these waves are indeed propagating in the azimuthal direction, we compute the real part of the Poynting vector to get the average power flow. Using equations (34.a) to (34.f), we get

$$\vec{S}_{av} = \frac{1}{2}Re[\vec{E}\times\vec{H}^*] = \frac{\omega\epsilon v a_0^2}{2x_{vn}^2\rho}\left|AJ_v\left(\frac{x_{vn}}{a_0}\rho\right)\cos\left(\frac{p\pi}{h}z\right)\right|^2 \hat{u}_\theta \quad (37)$$

This equation shows that the average power flow is in the azimuthal direction. It also shows that changing the sign of the azimuthal propagation constant ν reverses the direction of the power flow as expected.

Finally, the wave impedance of these waves in the azimuthal direction is given by

$$\frac{E_z}{H_\rho} = Z_{TM}(\omega) = \frac{x_{vn}^2\,\rho}{\omega\epsilon v a_0^2} \quad (38)$$

Not surprisingly, it depends on the curvature of the cavity through the radial distance ρ and the radius of the cavity $a_0$ and naturally on the frequency and the propagation constant

An important point that is not addressed in this paper is the issue of evanescence of azimuthally directed waves. Indeed, an examination of equation (23) shows that the root $x'_{vn}$ becomes imaginary when $\omega^2\epsilon\mu < \left(\frac{p\pi}{h}\right)^2$. The radial variation of the components of the electromagnetic field will then involve imaginary arguments with presumably no possibility of satisfying the boundary conditions at the walls of the cavity. The solution of the problem is based on the realization that the order of the Bessel function needed to satisfy the boundary condition itself becomes purely imaginary under these conditions. In other words, the azimuthal propagation constant ν becomes purely imaginary. However, when ν is imaginary, the components of the electromagnetic field as given by equations (25) become exponentially decaying or increasing as a function of θ. In other words, we have evanescent azimuthally directed waves. Naturally, similar conclusions are reached in regards to the TM modes and equation (32). Since these evanescent waves are not crucial to microwave components, and especially microwave dual-mode filters, they are not discussed any further here.

In summary, the observer in section III sees the waves described by equations (25) to (29) for TE modes and equations (34) to (38) for TM modes.

## V. RESULTS FOR TM MODES

In all what follows, we use an air-filled circular cavity of radius $a_0$=15mm and height h=45 mm. We first consider the TM modes.

### A. Propagation branches of TM modes

The azimuthal propagation constants of the TM modes are determined by solving equation (31) for successive values of ν. From equation (33), we know the starting point of each branch. The first ten roots of equation (33) are given in Table I. For each value of ν, equation (31) is solved numerically. Although not necessary, the range of ν is limited to the interval 0 to 10. Extension to a wider range poses no serious difficulty.

Table II First ten roots of $J_0(x)=0$

| 2.404826 | 5.520078 | 8.653728 | 11.79153 | 14.93092 |
|---|---|---|---|---|
| 18.07106 | 21.21163 | 24.352472 | 27.49348 | 30.63461 |

A plot of the roots $x_{vn}$ for the first ten branches, i.e., n=1 to 10 is shown in Figure 4. The lowest branch starts at ν=0 and $x_{01}$=2.404826. This is the first TM mode that starts propagating in the azimuthal direction in the uniform section of this circular cavity. The second branch starts at ν=0 and $x_{02}$=5.520078. Although the branches appear to be linear functions versus ν this is only approximately correct. However, for narrow-band systems such as dual-mode filters, a linear approximation may be sufficient.

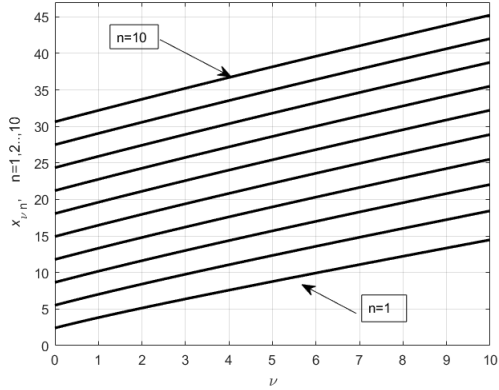

Figure 4. Roots of $J_\nu(x_{\nu n}) = 0$ vs "azimuthal propagation constant" $\nu$ for the first 10 branches. Each branch starts from a root of $J_0(x)=0$.

In order to understand why the modes are organized in branches, we examine the radial variation of the axial field ($E_z$) as function of $\rho/a$ with the azimuthal propagation constant $\nu$ as a parameter. Figure 5 shows the radial variation of $E_z$ for the first branch. The parameter $\nu$ varies from 0 to 10 in steps of 0.5. As this figure shows, for all values of $\nu$, the axial electric field vanishes exactly once in the interval $0<\rho\leq a$. This property is preserved over this entire branch. This is the reason these modes are organized in branches. The solutions of a branch evolve continuously from the seed solution at $\nu=0^+$. The case $\nu=0$ is an anomaly because of the behaviour for Bessel functions around the point $(\nu,x)=(0,0)$. We also note that as $\nu$ increases, the field is pushed towards the outer part of the cavity.

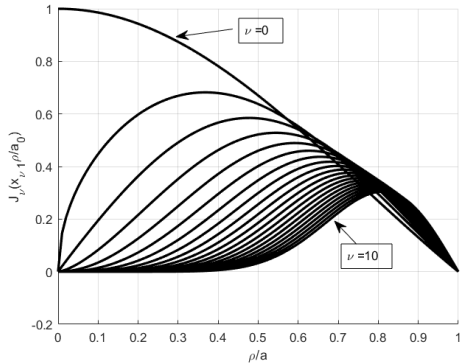

Figure 5. Radial variation of axial electric field of TM modes on the first branch.

To further show this property, figure 6 shows the radial variation of solutions in the second branch as function of $\rho/a$ for the same values of $\nu$ as in Figure 5. As clearly shown in this figure, for all values of $\nu$, the axial component $E_z$ vanishes *exactly twice* in the interval $0<\rho\leq a$. This is a common feature to all the solutions in this branch. In general, the axial component $E_z$ of the solutions of the $n^{th}$ branch will vanish exactly $n$ times in the interval $0<\rho\leq a$. Again the case of $\nu=0$ is an anomaly as it is the only solution that does not vanish at exactly $\rho=0$.

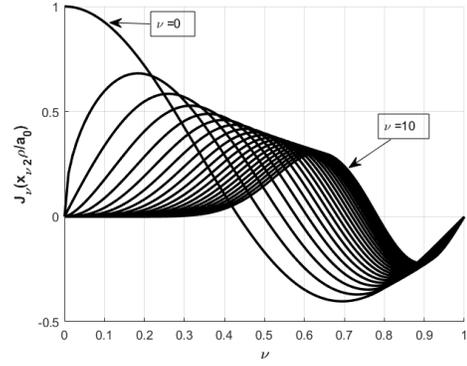

Figure 6. Radial variation of axial electric field of TM modes of the second branch.

### B. Dispersion relations of TM modes

The dispersion relation relates the azimuthal propagation constant $\nu$ to the frequency. This relationship is given by equation (32) for TM modes. Since we know by now how the roots $x_{\nu n}$ depend on $\nu$, this equation is used to express the frequency as ($c$ is the speed of light is vacuum)

$$f = \frac{c}{2\pi a_0}\sqrt{x_{\nu n}^2 + \left(\frac{p\pi a_0}{h}\right)^2} \quad (39)$$

With $a_0$=15mm, h = 45mm, this gives f in GHz as

$$f[GHz] = \frac{10}{\pi}\sqrt{x_{\nu n}^2 + \frac{p^2\pi^2}{9}} \quad (40)$$

For each value of $\nu$, the corresponding value of $x_{\nu n}$ is used in this equation to calculate the corresponding frequency. By following this procedure, the diagram in Figure 4 is transformed into a dispersion diagram. We examine two cases, $p$=0 and $p$=1. The resulting dispersion relations for the lowest three branches are shown in Figure 7 for $\nu$ smaller than 3. The modes are cut off at $\nu$=0. From this figure, we see that the first TM mode with $p$=0 and $p$=1 start propagating in the azimuthal direction at f=7.65 GHz and f=8.35 GHz, respectively for the given dimensions of the cavity.

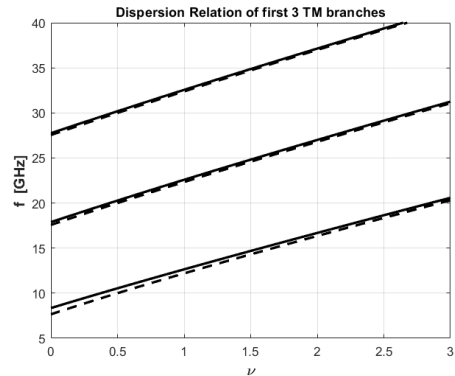

Figure 7. Dispersion curves of the three lowest TM branches for p=0 (dashed lines) and p=1 (solid lines). Air-filled cavity with $a_0$=15mm, h=45 mm.

## C. Resonant frequencies of TM modes

To keep the numbering of the resonances identical to the current nomenclature while allowing for arbitrary non-integer values of the azimuthal propagation constant, the modes will be denoted by TM$_{\nu np}$. The meaning of the indices $\nu$, n and p is as follows:
$\nu$ : azimuthal propagation constant which is also the order of the Bessel function. It is not necessarily an integer.
$n$ : branch number which also corresponds to the order of the root of the Bessel function of order $\nu$. The first branch corresponds to the first (smallest) root and so on.
$p$ : variation along the axis of the cavity (z-axis).

It is worth recalling that the boundary conditions in the radial direction and along the z-axis have already been imposed. The resonant frequencies will be determined once the boundary conditions in the azimuthal direction, i.e., the direction of propagation, are imposed. Indeed, the boundary conditions at $\theta=\theta_1$ and $\theta=\theta_2$, act as a sampling or filtering mechanism to select the discrete values at which independent or resonant solutions can exist. Once the required value of $\nu$ are determined, the values of the resonant frequencies are read out directly from Figure 7 for $p=0$ or $p=1$. Of course, it is always possible to solve equations (31) and (32) directly since Figure 7 is based on solving these equations. Since the procedure is the same for all branches, we examine only the lowest branch for different configurations. Only the cross section of the cavity is shown but it is assumed uniform in the axial direction. We first consider an empty cavity.

### 1) Empty circular cavity

In this case, the possible values of $\nu$ are obtained from the condition that the field be single valued as a function of the angle $\theta$. Since the complete range $0 \leq \theta \leq 2\pi$ is included in the domain of solution, $\nu$ must be an integer

$$\nu = m = 0,1,2,3 \ldots \quad (41)$$

The resonant frequencies are simply then given by equation (29), which is re-written as

$$f_{mnp} = \frac{c}{2\pi}\sqrt{\frac{x_{mn}^2}{a_0^2} + \left(\frac{p\pi}{h}\right)^2} \quad (42)$$

This is the well-known closed form expression of the resonant frequencies of TM modes in an air-filled cylindrical cavity of height $h$ and radius $a_0$ as given in any textbook on the subject [10, p. 290].

### 2) Circular cavity with perfectly conducting wedge

A perfectly conducting wedge with internal angle $\phi$ is inserted in the cavity as shown in inset of Figure 8. The height and radius of the cavity remain the same ($a_0$=15mm and h=45 mm). Waveguides with this type of cross sections were investigated in [11-13]. The possible values of the azimuthal propagation constant $\nu$ are obtained from the vanishing of the axial electric field at $\theta_1$=0 and $\theta_2$=2$\pi$-$\phi$. They are [13, p.23]

$$\nu = \frac{n\pi}{2\pi - \phi}, n = 1,2,3, \ldots \quad (43)$$

For each value of $\nu$, the correspond resonant frequency is read out from Figure 7 for $p=0$ without any further computation. The resonant frequencies of the TM modes of the lowest branch as a function of the wedge angle $\phi$ for n=1 are plotted in Figure 8 for p=0. For comparison, results obtained from HFSS are also shown. The agreement between the two results is excellent. The advantage of the method based on the propagation constant is evident. Whereas HFSS solves a new boundary problem for each value of the angle $\phi$, the method introduced in this paper simply computes the relevant values of $\nu$ from equation (43) and then reads out the resonant frequencies from Figure 7 without any further numerical computation. This is true for the resonant frequencies of each branch.

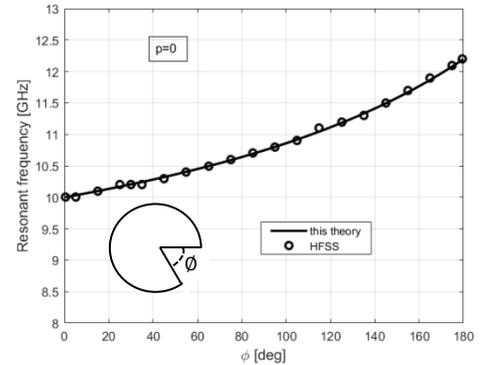

Figure 8. Resonant frequencies of lowest TM branch in a circular cavity with perfectly conducting wedge for $p=0$ versus the internal angle of the wedge ($a_0$=15 mm, h=45 mm).

## VI. RESULTS FOR TE MODES

### D. Propagation branches of TE modes

The azimuthal propagation constants of the TE modes are determined by numerically solving equation (22) for successive values of $\nu$. From equation (24), we know the starting point of each branch. The first ten roots of equation (24) are given in Table II. Although not necessary, the range of $\nu$ is limited to the interval 0 to 10. Extension to a wider range poses no serious difficulty.

Table II First ten roots of J$_1$(x)=0

| 0.0 | 3.831706 | 7.015587 | 10.17347 | 13.32369 |
|---|---|---|---|---|
| 16.47063 | 19.61586 | 22.76008 | 25.90367 | 29.04683 |

A plot of the roots $x_{\nu n}$ for the first ten branches, i.e., n=1 to 10 is shown in Figure 9. The lowest branch starts at $\nu$=0 and $x'_{01} = 0$. This is the first TE mode that starts propagating in the azimuthal direction in the uniform section of this circular cavity. This branch contains the dominant TE$_{11}$ mode. It is highly non-linear around the origin where it starts with an infinite slope. Given the importance of this branch, it will be discussed separately in mode details.

The second branch starts from $x'_{02} = 3.831706$ at ν=0. Although the branches look linear functions of ν, apart from the lowest branch, this is only approximately true. However, for narrow-band systems such as dual-mode filters, a linear approximation may be sufficient.

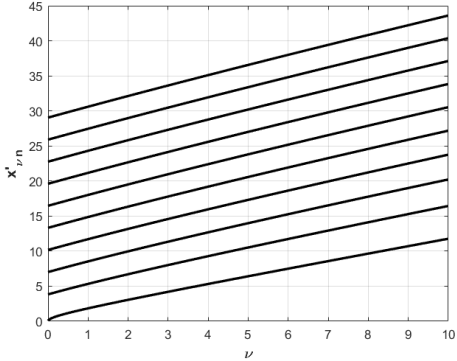

Figure 9. Roots of $J'_\nu(x'_{\nu n}) = 0$ vs "azimuthal propagation constant" ν for the first 10 branches. Each branch starts from a root of $J_1(x)=0$.

In order to understand why the modes are organized in branches, we examine the axial magnetic field (H$_z$) as a function of ρ/a with the azimuthal propagation constant as a parameter. Figure 10 shows radial variation of H$_z$ for the first TE branch. The parameter ν varies from 0 to 10 in steps of 0.5. As this figure shows, for all values of ν, the derivative of the axial field vanishes exactly once in the interval 0<ρ≤a. This property is preserved over this entire branch. This is the reason these modes are organized in branches. The solutions of a branch evolve continuously from the seed solution at ν=0$^+$.

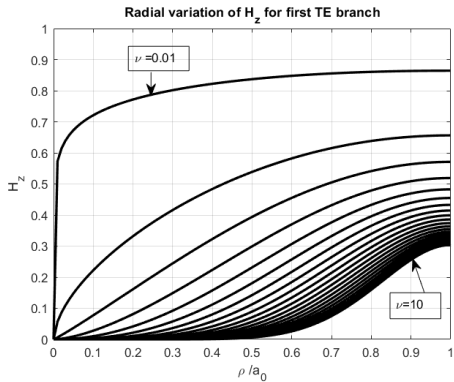

Figure 12. Radial variation of the axial magnetic field of the first TE branch for ν ranging from 0 to 10 in 0.5 steps.

To further show this property, figure 13 shows the radial variation of solutions in the second branch as function of ρ/a for the same values of ν as in Figure 12. As clearly shown in this figure, for all values of ν, the derivative of the axial component H$_z$ vanishes exactly twice (2) in the interval 0<ρ≤a. This is common feature to all the solutions in this branch. In general, the derivative of the axial component H$_z$ of the solutions of the n$^{th}$ branch will vanish exactly $n$ times in the interval 0<ρ≤a. Again the case of ν=0 is an anomaly as it is the only solution that does not vanish at exactly ρ=0.

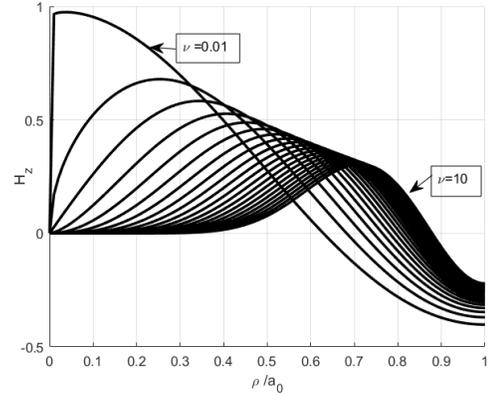

Figure 13. Radial variation of the axial magnetic field of the second TE branch for ν ranging from 0 to 10 in 0.5 steps.

*E. Dispersion relations of TE modes*

The dispersion relation relates the azimuthal propagation constant to the frequency. This relationship is given by equation (23) for TE modes. Since, we know by now how the roots $x'_{\nu n}$ depend on ν, this equation is used to express the frequency as

$$f = \frac{c}{2\pi a_0}\sqrt{(x'_{\nu n})^2 + \left(\frac{p\pi a_0}{h}\right)^2} \qquad (44)$$

With $a_0$=15mm, h = 45mm, this gives $f$ in GHz as

$$f[GHz] = \frac{10}{\pi}\sqrt{(x'_{\nu n})^2 + \frac{p^2\pi^2}{9}} \qquad (45)$$

For each value of ν, the corresponding value of $x'_{\nu n}$ is used in this equation to calculate the corresponding frequency. By following this procedure, the diagram in Figure 9 is transformed into a dispersion diagram. With $p$=1 in equation (45), the dispersion relations obtained from figure 9 are shown in Figure 14 for the three lowest TE branches. The fist TE mode with $p$=1 starts propagating in the azimuthal direction at $f$ = 3.333 GHz. This frequency is significantly lower than the cutoff frequency of the TE$_{11}$ mode (5.86 GHz).

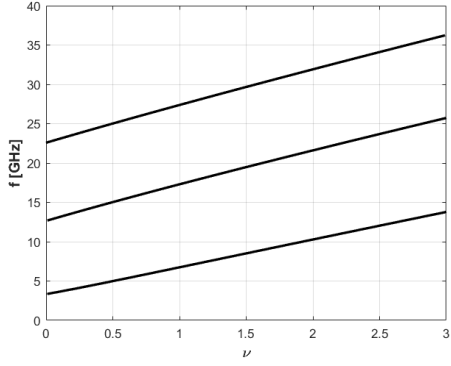

Figure 14. Dispersion of the first three TE branches for p=1 (Air-filled cavity with $a_0$=15mm, h=45 mm).

## F. Resonance frequencies of TE modes

Once the frequency dependence of the azimuthal propagation constant $\nu(\omega)$ is known, the resonant frequencies are determined by the boundary conditions on the angle $\theta$. These boundary conditions determine the values of $\nu$ at which resonance takes place. The corresponding resonant frequencies of the cavity are then simply read out from the dispersion diagram, or Figure 14.

We use the example of an empty circular cavity to illustrate the process. For an empty circular cavity, the requirement that the fields be single valued over the entire range of $\theta$, $0 \leq \theta \leq 2\pi$, we know that $\nu$ must be an integer

$$\nu = m = 0,1,2,3 \ldots \quad (46)$$

From equation (45), upon which Figure 14 is based, the resonant frequencies can be re-written as

$$f_{mnp} = \frac{c}{2\pi}\sqrt{\left(\frac{x'_{mn}}{a_0}\right)^2 + \left(\frac{p\pi}{h}\right)^2} \quad (47)$$

This is the well known closed form expression of the resonant frequencies of TE modes in a cylindrical cavity of height $h$ and radius $a_0$ as given in any textbook on the subject [10, p.290].

Given the importance of the lowest TE branch we examine its properties separately and in more depth in the following section.

## VII. THE LOWEST TE BRANCH

As we have seen earlier, the propagation constants of the azimuthally propagating TE modes are grouped in branches. Each root of the equation $J'_{\nu=0}(x') = -J_1(x') = 0$ determines the starting point (when $\nu$=0) of a branch. From the properties of Bessel functions, we know that this equation has a root at $x'$=0. It is its smallest (first) root and hence defines what we call the lowest TE branch. This root will be denoted by $x_{\nu 1}'$ with the number '1' standing for the first branch and $\nu$ the numerical value of the azimuthal propagation constant.

The propagation constant $\nu$ of the lowest branch is given by smallest root of the equation

$$J'_\nu(x'_{\nu 1}) = 0 \quad (48)$$

As explained earlier, it is more convenient to sweep over $\nu$ and determine the smallest root for each value of $\nu$ by solving equation (48). The result is given in Figure 15.

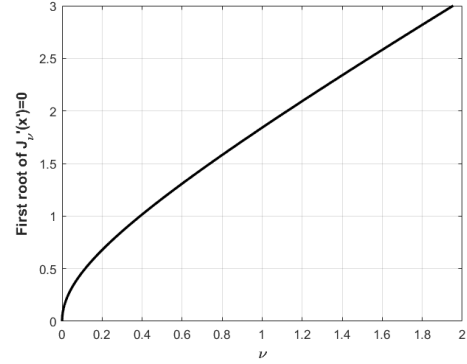

Figure 15. Smallest root $x_{\nu 1}'$ of $J'_\nu(x') = 0$ versus the azimuthal propagation constant $\nu$ of lowest TE mode.

As can be seen from this figure, the branch starts with infinite slope as $\nu$=0 and increases monotonically. It can be shown (see Appendix A for a derivation and a more accurate approximation) that for small values of $\nu$, we have

$$x'_{\nu 1} \sim \sqrt{2\nu}, \nu \to 0 \quad (49)$$

For the discussion that follows, we need the variation of $\nu$ versus frequency. Naturally this depends on the dimensions of the cavity. For $a_0$=15 mm and h=45 mm, it is obtained by using the smallest root of equation (48), for each value of $\nu$, in the following equation

$$f[GHz] = \frac{10}{\pi}\sqrt{x'^2_{\nu 1} + \frac{\pi^2}{9}} \quad (50)$$

This equation assumes that the variation of the field in the z-direction corresponds to $p$=1, the lowest possible value for TE modes. The dispersion relation obtained from these steps is shown in Figure 16.

The first thing to note about this dispersion relation is that it indicates that this mode starts propagating in the azimuthal direction at a frequency which is significantly lower than the cut-off frequency of the $TE_{11}$ mode in the circular waveguide. With the given dimensions the cut-off frequency of the $TE_{11}$ is

$$f_{cTE11} = \frac{1.841c}{2\pi a_0} = \frac{1.841 \times 3 \times 10^8}{2\pi \times 15 \times 10^{-3}} = 5.86\ GHz$$

However, Figure 16 shows that propagation in the azimuthal direction starts at f=3.33 GHz. Recall that propagation in the z-direction is already taking place since $k_z=\pi/h$. This result seems to contradict all current knowledge about propagation in empty cylindrical waveguides with circular cross section and needs closer scrutiny.

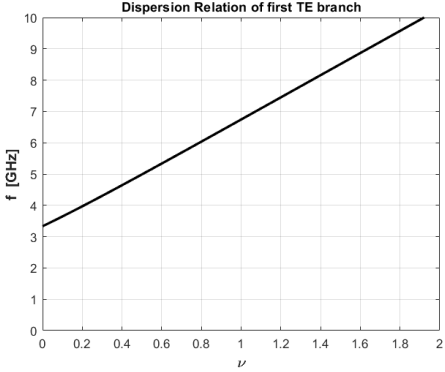

Figure 16. Dispersion relation of the lowest propagating TE branch for $0\leq\nu\leq 2$ for $a_0=15$ mm, $h=45$ mm and $p=1$.

The frequency at which the lowest azimuthally propagating mode starts propagating is obtained from equation (50) by setting $x'_{\nu 1}=0$. This gives

$$f_{c1} = \frac{10}{3} \text{ GHz} \quad (51)$$

This frequency is identical to the cut-off frequency of the dominant $TE_{10}$ in a rectangular cavity of cross section $a \times b = h \times a_0 = 45\text{mm} \times 15\text{mm}$. It is as if the mode sees this rectangular waveguide in the azimuthal direction. However, the curvature of the circular waveguide must have an effect on this mode. It turns out that this solution is too singular to be physical as we now show.

The axial component of the magnetic field is given by

$$H_z = A J_\nu\left(\frac{x'_{\nu 1}}{a_0}\rho\right)\sin\left(\frac{\pi z}{h}\right)e^{j\nu\theta} \quad (51)$$

The radial component of the electric field of this mode is given by

$$E_\rho = \frac{-j\omega\mu\nu}{x'^2_{\nu 1}\frac{\rho}{a_0}} A J_\nu\left(x'_{\nu 1}\frac{\rho}{a_0}\right)e^{j\nu\theta}\sin\left(\frac{\pi z}{h}\right) \quad (52)$$

Consider equation (49) and the fact that for small arguments, $J_\nu(x)\sim\left(\frac{x}{2}\right)^\nu$, we get

$$E_\rho \to \frac{-j\omega\mu}{2\frac{\rho}{a_0}} A \sin\left(\pi\frac{z}{h}\right), \nu \to 0 \quad (53)$$

Since the domain of solution includes the origin, $\rho=0$, this electric field stores an infinite amount of energy in the volume of the cavity. Therefore, it cannot represent a realizable physical solution for $\nu=0$. However, for any $\nu$ strictly positive, its energy is finite. It should therefore be possible to excite azimuthally propagating solutions for $\nu>0$. The challenge is to find ways to excite these solutions for $0<\nu<1$. The value $\nu=1$ is not included since this solution is already known to exist, it is the $TE_{111}$ resonance.

In order to get hints as to how these resonances might be generated, we use some facts from basics of wave propagation and transmission line theory. Consider the following problem:

A uniform medium supports an electromagnetic propagating mode with propagation constant $\beta(\omega)$ whose dependence on frequency is known. We assume that $\beta(\omega)$ is a monotonically increasing function of $\omega$, i.e., the medium does not support backward waves. A section of finite length, in the direction of propagation, is used to force resonances by placing terminations at the two ends of the line. The length of the line is allowed to vary from zero (0) to $l_{max}$, $0\leq l \leq l_{max}$. We would like to determine the frequency range of the resonances that can be set up on the line by only adjusting the terminations at the two ends. We identify four different cases.

*G. Full Wavelength resonance*

The first option is to set up resonances that appear when the length of the line is a multiple of the guided wavelength. This can be done by directly connecting to the two ends to form a ring. The lowest resonant frequency achieved by this configuration corresponds to

$$\beta(\omega_{min1}) = \frac{2\pi}{l_{max}} \quad (54)$$

The ring will then allow resonant frequencies in the range $\omega\geq\omega_{min1}$. It is important to keep in mind that a mathematical solution also exists at frequencies where the propagation constant vanishes as the TEM case. In the case of the lowest TE branch of the circular cavity, we have seen that this solution has an infinite energy and cannot be excited.

In the case of the circular cavity, condition (54) corresponds to an azimuthal wave running over the entire range of the angle $\theta$, i.e., $0\leq\theta\leq 2\pi$. From the requirement that the fields be single valued, we get the possible values of $\nu$ as $\nu=n$ where $n$ is an integer. With the value $n=0$ excluded, the smallest resonance frequency corresponds to $n=1$. This is nothing other than the $TE_{111}$ resonance of the cavity which can be read from the plot in Figure 16 as previously discussed.

*H. Half-Wavelength resonance ($0.5\leq\nu$)*

To force resonances at lower frequencies, the ends of the line are terminated by open circuits (magnetic walls) or short circuits (electric walls). At resonance, the length of the line is then a multiple of a half wavelength. The lowest resonant frequency achieved by this configuration corresponds to

$$\beta(\omega_{min2}) = \frac{\pi}{l_{max}} \quad (55)$$

Since we assumed that $\beta$ is a monotonically increasing function, we have $\omega_{min2}\leq\omega_{min1}$. The range of frequencies covered by this arrangement is $\omega_{min2}\leq\omega$.

The application of this arrangement to the circular cavity can be achieved by placing a perfectly conducting wedge inside the cavity as shown in Figure 17. The internal angle of the wedge is $\phi$. By applying the boundary condition, the possible values of $\nu$ are obtained [13]

$$\nu = \frac{n\pi}{2\pi-\phi}, n = 1,2, \quad (56)$$

Note that the value corresponding to n=0 is not taken into account as it results in the singular solution, or solutions in the higher branches. Excluding this singular case, the smallest value of ν corresponds to n=1. The corresponding values of ν as the internal angle of the wedge varies from 0 to 2π is ν≥0.5. With the resonance corresponding to ν=1 already set up in the previous case, we limit the values of ν to 0.5≤ν ≤1.

For the cavity with the given dimensions, the resonant frequencies can be read out from Figure 12 with no further calculations. The resonant frequency versus the internal angle of the wedge is shown in Figure 17. For comparison, results obtained from HFSS are shown as the circles. Very good agreement between the two results is achieved.

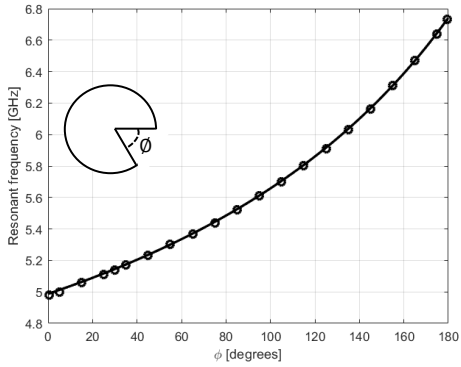

Figure 17. Resonant frequency of the first TE branch of a circular cavity with a perfectly conducting wedge versus its internal angle ϕ. (Air-filled cavity with $a_0$=15mm, $h$=45 mm, p=1). Circles are from HFSS.

This figure shows that we have propagation in both the azimuthal and the axial direction at frequencies down to about 5 GHz in this cavity. It is important to emphasize again that propagation is taking place in the azimuthal direction at frequencies that are lower than the cutoff frequency of the $TE_{11}$ mode (5.86 GHz).

*I. Quarter Wavelength resonance (0.25 ≤ν)*

To lower the resonant frequencies further, one end of the line is terminated in a short circuit (electric wall) and the other in an open circuit (magnetic wall). At resonance, the length of the line is then an odd multiple of a quarter wavelength. The lowest resonant frequency achieved by this configuration corresponds to

$$\beta(\omega_{min3}) = \frac{\pi}{2l_{max}} \quad (57)$$

Since we assumed that β is a monotonically increasing function, we have $\omega_{min3} \leq \omega_{min2}$. The range of frequencies covered by this arrangement is $\omega_{min3} \leq \omega$.

In order to implement this case in the circular cavity, we use a wedge with internal angle ϕ as in the previous case but with one face acting as a magnetic wall (open circuit). The second face of the wedge is perfectly conducting (short circuit).

Although the magnetic wall is not realizable over a wide frequency band, it provides a good numerical test of the theory.

The acceptable values of ν corresponding to this configuration can be determined by adapting the analysis in [11] resulting in (we ignore the singular case n=0)

$$\nu = \frac{n\pi}{2(2\pi-\phi)}, n = 1,2,\ldots \quad (58)$$

The smallest value of ν (other than zero) results when n=1 and ϕ=0. Since we have covered the range ν≥0.5, we limit the analysis to the range 0.25≤ν≤0.5. Once the values of ν are known, the corresponding resonant frequencies of the lowest TE branch for the cavity with the given dimensions are read out directly from Figure 16 without any further computation. These are shown in Figure 18 versus the internal angle of the wedge. Good agreement between the theory introduced in this paper and HFSS. However, whereas HFSS solves a new boundary-value problem for each value of ϕ, the propagation constant directly provides the correct resonant frequency for any value of ϕ.

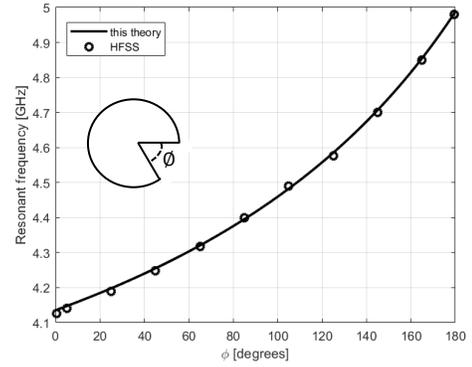

Figure 18. Resonant frequency of the first TE branch of a circular cavity with a wedge versus its internal angle ϕ. One face of the wedge is a perfect electric conductor (PEC) and the other a perfect magnetic conductor (PMC). (Air-filled cavity with $a_0$=15mm, $h$=45 mm, p=1). Circles are from HFSS.

As figure 18 shows, propagation in the azimuthal direction is taking place at frequencies as low as 4.136 GHz which is significantly lower than the cutoff of the $TE_{11}$ mode (5.86 GHz).

*J. Capacitively loaded shorted resonant line (ν≤0.25)*

This configuration is widely used in combline microwave filters to reduce the size of the combline resonators [16]. A gap between the top of the combline rod and the conducting walls acts as a capacitor. The energy stored in this capacitor requires a shorter line in order to balance the inductive energy as required for a resonance to be set up. If the normalized reactance of the capacitive loading is $-jx_c$, the resonance condition is

$$\beta(\omega_{min4}) = \frac{\tan^{-1}(x_c)}{l_{max}} \quad (59)$$

This equation shows that in the limit of $x_c \to 0$, the electric length of the line approaches zero at resonance. In particular,

if the propagation constant vanishes at zero, then a resonance as close to DC as desired can be achieved. Of course, there is a lower limit on the achievable values of $x_c$ as smaller values require larger capacitors.

To implement these conditions in the circular cavity, we use a wedge as in the previous two cases but force one face to have a capacitive surface impedance. The second face is still perfectly conducting. Unfortunately, the solution of this boundary value problem requires taking into account higher order modes, including TE and TM modes. This is beyond the scope of this paper. Instead, HFSS is used to determine the resonant frequency of the lowest TE resonance. The resonant frequency of the lowest TE mode as a function of the internal angle of the wedge is given in Figure 19 for a capacitive surface impedance $X_C=1\Omega$. It is again obvious that propagation is taking place in the azimuthal direction at frequencies that are significantly lower than the cutoff frequency of the $TE_{11}$ mode (5.86 GHz).

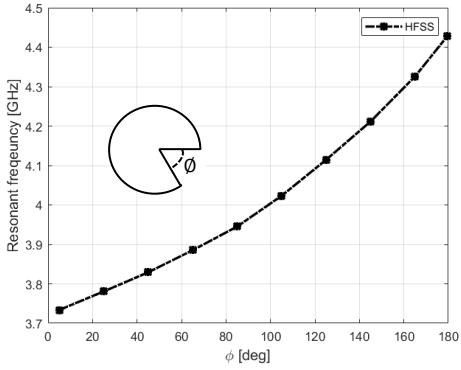

Figure 19. Resonant frequency of the first TE branch of a circular cavity with a wedge versus its internal angle ϕ as obtained from HFSS. One face of the wedge is a perfect electric conductor (PEC) and the other a normalized capacitive surface impedance $x_c$=-1. (Air-filled cavity with $a_0$=15mm, $h$=45 mm, p=1).

At this point one may be left to wonder as to whether there is a contradiction between our current knowledge and the results of the proposed theory in regards to propagation in circular waveguides and cavities. In fact, there is no contradiction. The modes of a circular cavity (or waveguide) are currently defined with the boundary conditions in the azimuthal direction *imposed* from the start. Consequently, the azimuthal propagation constant ν is forced to assume integer values for an empty cavity. These solutions are in fact *resonances* in the azimuthal direction. It is not uncommon for resonances to exist only at higher frequencies even when propagation of the wave takes place at much lower frequencies. For example, a quarter wave resonance on a TEM transmission line of finite length cannot happen below the frequency given by equation (57) despite the fact that a TEM wave propagates at all frequencies including DC. If we examine the azimuthal variation of the axial magnetic field of the $TE_{11}$ mode, equation (51) with ν=1, we see that it acquires a phase change of 360° around the circumference of the cavity at its cutoff frequency in the axial direction. The only way for this wave to acquire such a phase difference is to have a sizable propagation constant in the azimuthal direction at this frequency. This suggests strongly that it must start propagating in the azimuthal direction at a lower frequency. What the present theory does is tell us exactly at what frequency this wave starts propagating. In that sense, the theory introduced in this article complements our current knowledge by providing the missing information about the behavior of waves in circular cavities which has been limited to the information provided by resonance in the azimuthal direction.

VIII. DISCUSSION

Azimuthally propagating electromagnetic waves are only a special type of waves that can exist in a cylindrical cavity with circular cross section. Let us start with a source-free unbounded uniform and isotropic medium with constant parameters μ and ε. A cylindrical system of coordinates (ρ,θ,z) is used.

The non-zero z-component of the electromagnetic field of an elementary propagating wave in such a medium takes the general form [15, pp. 198-202]

$$\psi_{k_c,\nu,k_z} = e^{\pm j\nu\theta}e^{\pm jk_z z}\begin{cases}H_\nu^{(1)}(k_c\rho)\\H_\nu^{(2)}(k_c\rho)\end{cases} \quad (60.a)$$

$$k_c^2 + k_z^2 = \omega^2\mu\varepsilon \quad (60.b)$$

Here, $H_\nu^{(1)}$ and $H_\nu^{(2)}$ are Hankel functions of the first and second kind respectively. These two functions describe propagation in the radial direction. Equation (60.a) is a local solution to Helmholtz equation representing waves that can propagate in any direction in the unbounded medium. If boundary conditions are imposed the set of solutions is reduced in different ways.

Since the problem is linear, a linear superposition of elementary waves is also a solution. By using different combinations, we can force the wave to propagate in different directions. Let us consider the following combination

$$\psi_1 = e^{-j\nu\theta}e^{-jk_z z}\frac{1}{2}\left[H_\nu^{(1)}(k_c\rho) + H_\nu^{(2)}(k_c\rho)\right]$$
$$= J_\nu(k_c\rho)e^{-j\nu\theta}e^{-jk_z z} \quad (61)$$

We see that this wave is no longer propagating in the radial direction since $J_\nu(k_c\rho)$ is real for real arguments and order. If we include the time in equation (60), we get

$$\psi_1(\rho,\theta,z,t) = J_\nu(k_c\rho)\cos(\omega t - \nu\theta - k_z z) \quad (62)$$

It is obvious that this wave is propagating in both the azimuthal and the axial directions at the same time.
Let us now consider the following combination

$$\psi_2 = J_\nu(k_c\rho)e^{-jk_z z}\frac{1}{2}\left[e^{-j\nu\theta} + e^{j\nu\theta}\right]$$
$$= J_\nu(k_c\rho)\cos(\nu\theta)e^{-jk_z z} \quad (63)$$

If we include the time, this wave becomes

$$\psi_2(\rho,\theta,z,t) = J_\nu(k_c\rho)\cos(\nu\theta)\cos(\omega t - k_z z) \quad (64)$$

It is evident that this wave is propagating only in the z-direction. It is a linearly polarized wave since its maximum as a function of the angle θ does not change with time. It describes plane waves in the sense that the equiphase surfaces are planes orthogonal to the z-axis. If the domain of solution includes the complete range 0≤θ≤2π, we know that ν assumes only integer values. This type of waves are the standard solutions that are presented in textbooks about propagating modes in circular waveguides.

Finally consider the following combination

$$\psi_3 = J_\nu(k_c\rho)e^{-j\nu\theta}\frac{1}{2}[e^{-jk_z z} + e^{jk_z z}]$$
$$= J_\nu(k_c\rho)\cos(k_z z)e^{-j\nu\theta} \quad (65)$$

If we include the time, this wave becomes

$$\psi_3(\rho,\theta,z,t) = J_\nu(k_c\rho)\cos(k_z z)\cos(\omega t - \nu\theta) \quad (66)$$

It is evident that this wave propagates in the azimuthal direction. Even if the value of ν is restricted to be an integer, this function represents a wave that is propagating in the azimuthal direction. It is straightforward to set up and observe these rotating waves in a cylindrical cavity by using a full-wave solver such as HFSS. Furthermore, for ν≠0, it represents circularly polarized waves. Since a circularly polarized wave can be decomposed into two linearly polarized waves, it represents two degenerate modes for ν≠0. In particular, for ν=1 which corresponds to the TE$_{11}$ mode of a circular cavity, the two degenerate polarizations are intrinsically tied together in this azimuthally propagating wave. In this sense, azimuthally propagating waves are the natural setting for phenomena that involve both TE$_{11}$ polarizations simultaneously. This is obviously the case of dual-mode filters in circular cavities. It will be demonstrated in future articles that dual-mode filters that outperform the state-of-the-art are possible thanks to this characteristic.

One might argue that the parameter ν can assume only integer values in uniform cavities thereby limiting the need for the in-depth analysis presented in this article. In real world microwave filter applications, such as dual-mode filters, uniform cavities are never used. To control the signal and design this class of components discontinuities such as tuning, coupling and polarizing elements are inserted in the cavity. The resulting perturbed cavity is no longer uniform and cannot support the modes of the uniform cavity without coupling them. However, the perturbed cavity contains uniform sections between the perturbations where azimuthal waves propagate. The cavity is then accurately represented by transmission line sections in the azimuthal directions which are separated by the perturbations. For microwave filters based on the degenerate TE$_{11}$ modes, the transmission line sections involve values of ν around ν=1. It becomes then crucial to know how ν varies with frequency in order to model the frequency response of the filter. The theory presented in this paper provides the required frequency dependence.

## IX. APPLICATION TO DUAL-MODE FILTERS

The original goal of this study was to develop models of microwave filters in circular and squircle cavities based on propagation and not resonance. To that end, the variation of the propagation constant of azimuthally propagating modes as a function of frequency is required. As argued in earlier sections, when applicable, propagation-based models contain significantly more information than resonance-based models. Fortunately, robust microwave filter designs can be carried out using cavities in which modes can propagate in uniform sections that are separated by discontinuities. We give a simple illustrative example here, more complex designs will be presented in future articles. Even though the appearance of the lowest branch in the TE mode is arguably the most surprising result of this study, actual applications involve the part of the branch that is located around ν=1. In that sense, the fact that solutions around ν=0 do not seem to be accessible, is of little important to practical applications such as dual-mode filters

*Design example.*

We consider a dual-mode circular cavity as shown in Figure 20. The input and output are coaxial probes that are placed 90° apart. Perturbations that extend over the total height of the cavity are placed midway between the probes at 135°. The radius and the height of the cavity are 15 mm and 45 mm respectively. With these dimensions, the resonant frequency of the degenerate TE$_{111}$ can be read directly from Figure 9 resulting in f$_r$ = 6.74 GHz.

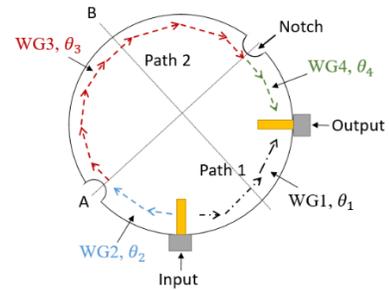

Figure 20. Top view of a circular waveguide cavity with coaxial input/output and longitudinal grooves.

As explained in above, energy between the input and output of the dual-mode cavity propagates azimuthally in opposite directions. The addition of the electromagnetic waves propagating in opposing directions can be added together to create a second order passband at the branch that is located around ν=1 or $\nu = m = 1,2,3$ whereas the addition of the electromagnetic waves propagating in opposing directions can cancel out to create two real-frequency transmission zeros at the same frequency or two different of frequencies. This phenomenon is not limited to TE branch of circular cavities. In fact, azimuthally propagating waves exist in squircle, and waveguide loaded cavities [waveguide loaded dual-mode Bakr].

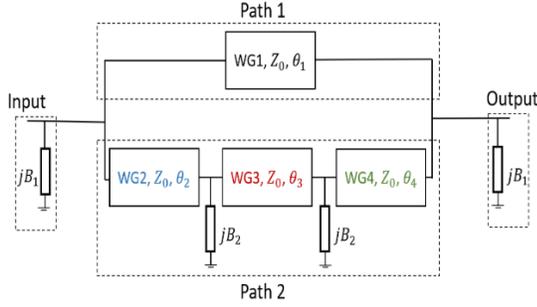

Figure 21. Equivalent circuit based on propagating waves in the azimuthal direction.

It is argued that microwave filters would be better modelled and designed based on the concept of propagation and not resonance. The equivalent model shown in Figure 21 is based azimuthally propagating waves in opposing directions. The model includes a wealth of information such as the dominant physics of the problem, the polarisation of the travelling waves, symmetry conditions, propagation constant and wave impedance as a function of frequency, and all perturbations in place. The dispersion of the TE branch through the $TE_{111}$ resonance is linear and is corresponding to

$$\theta^0 = \theta^r \frac{f + f_{rTE211} - 2f_{rTE11}}{f_{rTE211} - f_{rTE111}} \qquad (67)$$

Here, $f_{rTE111}$ and $f_{rTE111}$ are the resonant frequencies of the $TE_{211}$ and $TE_{111}$. The approximation is invalid for the lowest branch in the TE mode. For dual-mode filters, the dispersion of the lowest TE mode is too far away in frequency and can be ignored. The application of the lowest TE mode will be detailed in future articles. The wave impedance in the azimuthal direction depends on the distance $1/\rho$. For narrowband applications, the frequency dependence of the wave impedance is negligible and will be ignored in this example. The normalised circuit parameters of the branch line model in Fig. 21 are $Z_0 = 1, B_1 = 0.0242, B_2 = 0.0096$. The parameter $B_1$ is preceded and followed by a transformer at the input and the output with normalised impedance of 4.9751 ohm.

A comparison between the response of the equivalent circuit and the HFSS model is shown in Figure 22. The solid lines correspond to the HFSS model, and the dotted lines correspond to the propagation-based model. The agreement is excellent. As argued earlier, the model is predicating the appearance of a second order bandpass filter and two real-frequency transmission zeros due to coherent combination and cancellation of propagating waves in opposing directions. The model goes further to show the dependence of transmission zeros on the nature and location of perturbations. Here, the grooves are of inductive nature. As seen in Figure 22, the two transmission zeros are located on the imaginary axis of the s-plane when the grooves are at A whereas the transmission zeros become complex when the grooves are at B. The dispersion of the coupling elements and perturbations cause the weak asymmetry in the frequency response. The structure can be cascaded to obtain higher-order filters having arbitrary N real-frequency transmission zeros. The design methodology and physical dimensioning will be detailed in future papers.

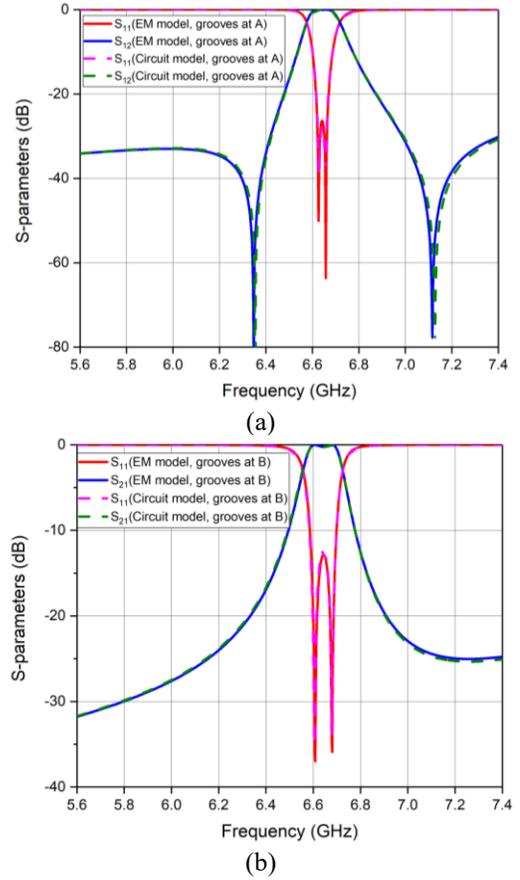

Figure 22. Comparison between the scattering response of the propagation-based mode and EM model with grooves as location A and B (Fig. 21).

X. CONCLUSIONS

A theory of azimuthally propagating electromagnetic waves in a cylindrical cavity with circular cross section was introduced. It is reported that the lowest TE mode starts propagating in the azimuthal direction at a frequency that is determined only by the height of the cavity. This cut-off frequency can be significantly lower than the cut-off frequency of the $TE_{11}$ mode in the axial direction. A natural organization of the modes emerges. Each root of the Bessel function $J_0(x)$ determines one branch of the TM modes. Similarly each root of the Bessel function $J_1(x)$ determines one branch of the TE modes. The lowest TE branch starts from a non-physical resonance at $v=0$ and $x'=0$. The variation of the azimuthal propagation constant with frequency was derived for both TE and TM modes. Below cut-off, the azimuthal propagation constant $v$ becomes purely imaginary. This dispersion relation plays a major role is developing models of dual-mode filters that are based on azimuthally propagating waves. A dual-mode filter of order 4, including measured results, was used to demonstrate the suitability of models based on azimuthally propagating waves.

## References

[1] S. B. Cohn, "Direct-Coupled-Resonator Filters," in *Proceedings of the IRE*, vol. 45, no. 2, pp. 187-196, Feb. 1957.
[2] P. I. Richards, "Resistor-Transmission-Line Circuits," in *Proceedings of the IRE*, vol. 36, no. 2, pp. 217-220, Feb. 1948.
[3] R. J. Wenzel, "Synthesis of Combline and Capacitively Loaded Interdigital Bandpass Filters of Arbitrary Bandwidth," in *IEEE Transactions on Microwave Theory and Techniques*, vol. 19, no. 8, pp. 678-686, Aug. 1971.
[4] J. C. Slater, Microwave Electronics, New York, Van Nostrand, 1950.
[5] S. A. Schelkunoff, "On representation of electromagnetic fields in cavities in terms of natural modes of oscillation", *J. Appl. Phys.*, vol. 24, pp. 262-267, March 1953.
[6] K. Kurokawa, "The Expansions of Electromagnetic Fields in Cavities," in *IRE Transactions on Microwave Theory and Techniques*, vol. 6, no. 2, pp. 178-187, April 1958.
[7] G. Reiter, "Solution of field equations for strongly coupled cavity systems" in Electromagnetic Wave Theory, New York, Pergamon, pp. 357-367, 1965.
[8] M. S. Bakr, and I. C. Hunter, " Resonator apparatus and method of use thereof," Patent GB2567962A, May 01, 2019.
[9] M. S. Bakr, I. C. Hunter and W. Bosch, "Broadband Dual-Mode Dielectric Resonator Filters," *2018 International Conference on Microwave and Millimeter Wave Technology (ICMMT)*, 2018, pp. 1-3, doi: 10.1109/ICMMT.2018.8563323.
[10] D. M. Pozar, Microwave Engineering, Wiley, New York, 2012.
[11] S. A. Schelkunoff, Electromagnetic Fields, Blaisdell, New York, 1963.
[12] I. S. Gradshteyn and I. M Ryzhik, Tables of Integrals Series and Products, Academic Press, 1994.
[13] R. E. Collin, Field Theory of Guided Waves, IEEE Press, 1991.
[14] S. A. Schelkunoff, Electromagnetic Waves, Van Nostrand, New York 1943.
[15] A Elsherbini, D. Kajfez and S. Zeng, "Circular Sectoral Waveguides," IEEE Antennas and Propagation Magazine, vol. 33, No.6, pp. 20-27, Dec. 1991.
[16] I. Hunter, Theory and Design of Microwave Filters, IET, 2001.
[17] R. Harrington, Time Harmonic Electromagnetic Fields, McGraw-Hill, New York, 1961.

## Appendix A

This appendix gives the variation of the smallest root $x'_{\nu 1}$ of $J'_\nu(x) = 0$ when the order $\nu$ of the Bessel function approaches zero.

The derivative of the Bessel function is given by

$$J'_\nu(x) = \frac{\nu}{x} J_\nu(x) - J_{\nu+1}(x) \quad (A.1)$$

For small values of $x$, we have

$$J_\nu(x) \sim \frac{1}{\Gamma(\nu+1)} \left(\frac{x}{2}\right)^\nu, \quad x \to 0 \quad (A.2)$$

Using the identity of the gamma function $\Gamma(\nu + 1) = \nu \Gamma(\nu)$ in the combination of (A.1) and (A.2), the smallest root $x'_{\nu 1}$ of $J'_\nu(x) = 0$ satisfies

$$\frac{1}{\Gamma(\nu+1)} \left(\frac{x'_{\nu 1}}{2}\right)^\nu \left(\frac{\nu}{x'_{\nu 1}} - \frac{x'_{\nu 1}}{2}\right) = 0 \quad (A.3)$$

From this equation, we get

$$x'_{\nu 1} \sim \sqrt{2\nu}, \quad \nu \to 0 \quad (A.4)$$

Using this result in equation (42), we get the dispersion relation of the lowest TE mode in the limit of small values of $\nu$

$$f \sim \frac{cp}{2h} + \frac{ch}{2p\pi^2 a_0^2} \nu, \quad \nu \to 0 \quad (A.5)$$

We see that in this limit, the propagation constant of the waves of this branch is a linear function of the frequency.

A more accurate approximation results when second order terms are kept in the expansion of the Bessel functions in (A.1). The result is

$$x'_{\nu 1} \sim 2(1+\nu)\sqrt{\frac{\nu}{2+3\nu}}, \quad \nu \to 0 \quad (A.6)$$

The corresponding dispersion relation is given by

$$f \sim \frac{cp}{2h} + \frac{ch}{p\pi^2 a_0^2} \frac{\nu(1+\nu)^3}{2+3\nu}, \quad \nu \to 0 \quad (A.7)$$